# Magnetic field inside the metal induced by anisotropic electronic pressure

I. V. Oladyshkin, D. A. Fadeev, V. A. Mironov

Institute of Applied Physics of the Russian Academy of Sciences, 603950, Nizhny Novgorod, Russia

oladyshkin@gmail.com

We show theoretically that anisotropy of electronic distribution function inside the laser-irradiated metal leads to the formation of edge currents at the timescale of distribution isotropization. When the electronic pressure in the skin-layer is anisotropic, pressure gradient appears to be non-potential force effectively producing low-frequency magnetic field. In typical experiments with femtosecond laser pumping generated internal magnetic field can rich magnitude up to ~1 Tesla in non-damaging interaction regime. We demonstrate that this field is localized inside the metal, while just a minor part of its energy can be radiated into free space as a sub-terahertz signal.

Interaction of femtosecond optical pulses with solids is usually accompanied by a variety of ultrafast phenomena like nonequilibrium heating, ballistic heat transfer, crystal lattice destruction, generation of optical harmonics and terahertz signal. In all of the listed effects dynamics of electron distribution function plays a key role. However, subpicosecond kinetics of charge carriers in solids is still a relevant problem both due to high computational complexity of numerical modeling and technical limitations of ultrafast measurements.

For example, dynamics of distribution function isotropization and corresponding elementary processes in nonequilibrium metals are poorly represented in relevant experimental and theoretical literature. In most of theoretical papers the authors study the relaxation of electronic energy supposing that the velocity distribution is isotropic [1-4]. There are also experiments investigating the energy distribution evolution [5-9], however it is difficult to find any direct observations of anisotropy relaxation.

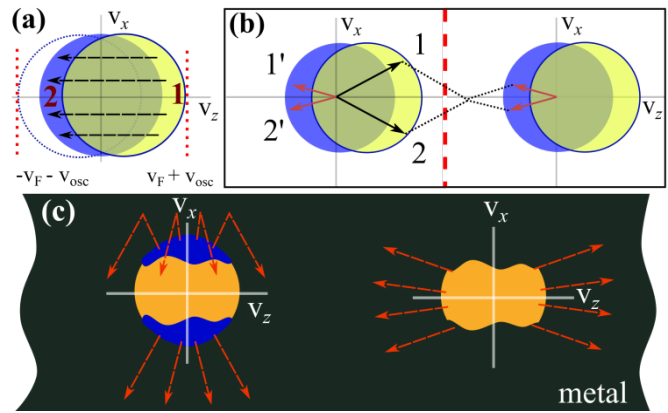

**Fig. 1** Possible mechanisms of anisotropic distribution function formation. a) Illustration of the detailed balance principle: for the shifted distribution function electron scattering from "region 1" to "region 2" is the most probable process. b) An example of Umklapp collision increasing the distribution anisotropy: the sum of electron momentum vectors 1 and 2 ends up in the neighboring Brillouin zone, so the scattering to 1' and 2' is possible. c) Schematic picture, illustrating fast loss of electrons moving perpendicularly to the metal surface.

At the same time, there are several physical mechanisms which may produce or keep the anisotropy of electronic distribution function. First of all, formation of anisotropic distribution function under the action of linearly polarized oscillating field is a consequence of the detailed balance principle combined with the Pauli blocking effect (see Fig. 1, a). Other possible mechanisms are Umklapp processes, ballistic transport and phonon drag effect. The first one is important if the Fermi surface covers rather small part of the Brillouin zone: in that case Umplapp processes leads to the effective backward scattering of electrons (Fig. 1, b).

Under the ballistic transport effect we mean the case when the average electron free path is comparable or larger than the optical skin depth. Hot electrons moving perpendicularly to the surface leave the skin layer during 10-15 fs while the other part of electrons keep moving roughly parallel to the border. Because the particles returning from the bulk have lower average energy, the anisotropic velocity distribution should be formed due to the described process (see Fig. 1, c).

So-called phonon drag effect may also keep the distribution anisotropy. Total momentum of free electrons and crystal lattice inside the skin-layer may change only due to phonon-phonon scattering or momentum transport perpendicular to the surface. But such processes have significant delay (depending on the material), and some phonons propagating parallel to the surface will return their momentum to the electrons before leaving the active region or before transformation into the different oscillating mode. This may significantly prolong the isotropization process [10].

In this Letter we focus only on the electrodynamical phenomena caused by the anisotropic velocity distribution of electrons. The microscopic origination of anisotropy depends on concrete metal and (mostly) on its temperature, while the electrodynamical features can be described in the general case. In the first part of the Letter we propose the simplest analytical description using modified hydrodynamic equations derived from the Boltzmann equation for electron gas. We demonstrate that anisotropic electronic pressure produced in the optical skin-layer of metal appeared to be non-potential external force (with rot ≠ 0) exciting electric currents with nonzero vorticity. As a consequence, quite a strong magnetic field inside the metal is generated with a characteristic timescale of the pressure isotropization. We also show that after the excitation this field propagates inside the metal in accordance with the diffusion-like equation. All of the obtained analytical results are illustrated by numerical modeling in the second part of the paper.

We believe that besides general interest to laser-metal interaction physics, developed theory can be useful for interpretation of experiments on optical-to-THz conversion, second harmonic generation and ultrafast demagnetization of metals. Below we use the example of THz response of metals since this effect is probably most closely related to the discussed electromagnetic phenomena.

Consider the action of femtosecond laser pulse obliquely incident onto the metal. Due to large permittivity of metals, electric field of optical wave inside the metal is mostly parallel to the surface for any polarizations and incidence angles. For clarity, here we discuss only *p*-polarized light because it is optimal for optical-to-THz conversion [11-13] (however, consideration is also valid for any other polarizations). The schematic picture of evolution of electronic distribution function is shown in Fig. 2.

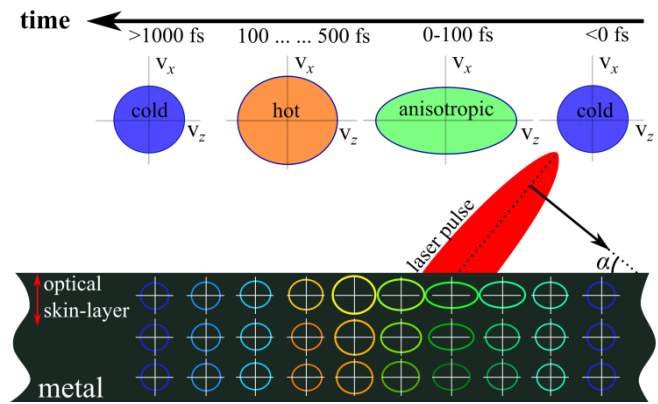

**Fig 2**. Schematic picture of electronic distribution function evolution in time (top) and in space (bottom).

For clarity, suppose that the laser beam diameter is much bigger than its length; this assumption is valid for most of non-destructive experiments with millijoule pumping, when the typical beam diameter is 1-10 mm and the pulse duration is 50-100 fs which corresponds to the length of 15-30 microns [11-13]. In this case the interaction area moves along the metal surface with the velocity c/cosα (see Fig. 2) and the problem is effectively two-dimensional.

Let us consider the distribution function of free electrons $f(x, z, v_x, v_z, t)$ and find the self-consistent system of equations on low-frequency electromagnetic fields inside the metal. Further the term "low-frequency" is used for the processes with a characteristic timescale of the order of laser pulse envelope, or distribution function isotropization, or longer (in contrast to the oscillations at optical frequency). In the analytical description we use perturbation theory, supposing that the momentum distribution is mostly determined by the external optical pumping and scattering processes, while relatively weak internal low-frequency fields influence only on slow spatial redistribution of electrons. We start from the Boltzmann equation:

$$\frac{\partial f}{\partial t} + \mathbf{v}\frac{\partial f}{\partial \mathbf{r}} - \frac{e\mathbf{E}}{m}\frac{\partial f}{\partial \mathbf{v}} = \left(\frac{\partial f}{\partial t}\right)_{col}, \tag{1}$$

where $\mathbf{E}$ is the self-consisted low-frequency electric field, $e$ and $m$ are the electron charge and mass respectively and the term in the right side describes the influence of collision processes.

Let us derive the Euler-type equation form Eq. (1) multiplying it by the velocity $\mathbf{v}$ and integrating over the velocity space:

$$\frac{\partial}{\partial t}\int \mathbf{v} f d\mathbf{v} + \int \mathbf{v}\left(v_x\frac{\partial f}{\partial x} + v_z\frac{\partial f}{\partial z}\right) d\mathbf{v} - \int \mathbf{v}\left(\frac{eE_x}{m}\frac{\partial f}{\partial v_x} + \frac{eE_z}{m}\frac{\partial f}{\partial v_z}\right) d\mathbf{v} = \int \mathbf{v}\left(\frac{\partial f}{\partial t}\right)_{col} d\mathbf{v}. \tag{2}$$

We can neglect the terms like $\int v_x v_z \frac{\partial f}{\partial x} d\mathbf{v}$ since they are of the second perturbation order (proportional to the low-frequency current squared). After the integration we find:

$$\frac{\partial}{\partial t} n\langle \mathbf{v}\rangle + \mathbf{x}_0 \frac{\partial}{\partial x} n\langle v_x^2\rangle + \mathbf{z}_0 \frac{\partial}{\partial z} n\langle v_z^2\rangle - \frac{ne\mathbf{E}}{m} = -\nu n\langle \mathbf{v}\rangle, \tag{3}$$

where $\mathbf{x}_0$ and $\mathbf{z}_0$ are the unit vectors, $n$ is the electron density, $\langle \mathbf{v}\rangle$ is the average electron velocity, $\langle v_{x,z}^2\rangle$ are the average squared velocities in $x$ and $z$ directions respectively and $\nu$ is the effective transport collision frequency. Below we use $\mathbf{v}$ instead of $<\mathbf{v}>$, because the hydrodynamical velocity is an averaged characteristic in itself. We also denote $\langle v_{x,z}^2\rangle$ as $T_{x,z}/m$, introducing two "temperatures" $T_{x,z}$ along the $x$ and $z$ directions.

Integration of the Boltzmann equation (1) over the velocity space gives the standard continuity equation:

$$\frac{\partial n}{\partial t} + div\, n\mathbf{v} = 0. \tag{4}$$

Now let us find the curl (*rot* in our definition) of the Eq. (3):

$$n_0 \frac{\partial}{\partial t} rot\, \mathbf{v} = -\frac{n_0}{m}\frac{\partial^2 (T_x - T_z)}{\partial x \partial z} - \frac{n_0 e}{m} rot\, \mathbf{E} - \nu n_0\, rot\, \mathbf{v}. \tag{5}$$

For the simplicity of derivation we consider the case of relatively high collision frequency $\nu \gg \frac{\partial}{\partial t}$. However, it is not a principal limitation and it is not used in our numerical modeling (see below). In this way we find

$$rot\, \mathbf{v} \cong \frac{1}{m\nu}\frac{\partial^2 (T_z - T_x)}{\partial x \partial z} - \frac{e}{m\nu} rot\, \mathbf{E}. \tag{6}$$

To complete the system, one needs Maxwell equations:

$$rot\ \mathbf{E} = -\frac{1}{c}\frac{\partial \mathbf{H}}{\partial t}, \tag{7}$$

$$rot\ \mathbf{H} = \frac{1}{c}\frac{\partial \mathbf{E}}{\partial t} - \frac{4\pi}{c}ne\mathbf{v}. \tag{8}$$

Eqns. (7) and (8) give:

$$\Delta\ \mathbf{H} - \frac{1}{c^2}\frac{\partial^2 \mathbf{H}}{\partial t^2} = \frac{4\pi n_0 e}{c} rot\ \mathbf{v}. \tag{9}$$

Now substituting rot $\mathbf{v}$ from the Eq. (6) we find the equation describing magnetic field excitation by non-potential part of the electronic pressure:

$$\Delta\ H_y - \frac{1}{c^2}\frac{\partial^2 H_y}{\partial t^2} - \frac{\omega_p^2}{c\nu}\left[\frac{1}{e}\frac{\partial^2 (T_z - T_x)}{\partial x \partial z} + \frac{1}{c}\frac{\partial H_y}{\partial t}\right] = 0. \tag{10}$$

where $\omega_p^2 = 4\pi n_0 e^2/m$ is the plasma frequency in metal. Since the optical skin-layer depth is much smaller than the laser pulse length, we use the following inequalities:

$$\frac{\partial^2 H_y}{\partial x^2} \gg \frac{\partial^2 H_y}{\partial z^2}, \frac{1}{c^2}\frac{\partial^2 H_y}{\partial t^2}$$

Under this assumption Eq. (10) becomes the equation of magnetic field diffusion with a source, proportional to the difference of "temperatures" (i.e. to the distribution function anisotropy):

$$\frac{\partial H_y}{\partial t} - \frac{c^2 \nu}{\omega_p^2}\frac{\partial^2 H_y}{\partial x^2} = -\frac{c}{e}\frac{\partial^2 (T_z - T_x)}{\partial x \partial z}. \tag{11}$$

Here we can introduce magnetic diffusivity $D_H = c^2\nu/\omega_p^2$ which is typically from 100 to 1000 cm$^2$/sec in metals like Au, Cu, Al, Ag and others. So, $D_H$ is of the same order as the electron thermal diffusivity $D_T$ in these metals, but $D_T$ depends on the collision frequency inversely. According to Eq. (11), diffusion of magnetic field from the initially heated skin-layer $l_{sk} = c/\omega_p$ takes $1/\nu$ sec, which is typically 5-30 fs depending on metal and its temperature. Note that similar diffusion-like equation for the magnetic field was obtained in the theoretical study of photon-drag effect inside metals [14].

From Eq. (11) it is easy to estimate maximal magnetic field $H_y$ for the case of short laser pulse and slow magnetic diffusion:

$$H_{y\ max} \cong \frac{\cos\alpha}{e}\frac{\partial (T_z - T_x)}{\partial x}. \tag{12}$$

For the temperatures' difference 1 eV and the skin-layer depth 10 nm we obtain $H_{y\ max} \cong 0.5$ Tesla. Note that typical ablation thresholds of metals correspond to maximal electronic temperatures of about 10 eV and higher.

Now let us describe the distribution of electric field. Divergence of Eq. (3) gives us an equation on the perturbation of electron density:

$$\frac{\partial^2 n}{\partial t^2} + \nu\frac{\partial n}{\partial t} - \frac{\partial^2}{\partial x^2}\frac{nT_x}{m} - \frac{\partial^2}{\partial z^2}\frac{nT_z}{m} + \omega_p^2 \delta n = 0. \tag{13}$$

Since the plasma frequency is very high ($\sim 10^{16}$ s$^{-1}$), Eq. (13) has a trivial quasistatic solution similar to once obtained in [15] for the case of thermalized electrons:

$$\delta n \cong \frac{1}{4\pi e^2}\frac{\partial^2 T_x}{\partial x^2}. \tag{14}$$

Then, using Eq. (14), Maxwell equations and the inequality $\frac{\partial^2}{\partial z^2} \ll \frac{\partial^2}{\partial x^2}$ we find the following relation:

$$E_z = -\frac{1}{e}\frac{\partial T_z}{\partial z} + \frac{D_H}{c}\frac{\partial H_y}{\partial x}. \tag{15}$$

Finally, diffusion equation (11) can be rewritten for the electric field using the above relation:

$$\frac{\partial E_z}{\partial t} - D_H \frac{\partial^2 E_z}{\partial x^2} = -\frac{1}{e}\frac{\partial^2 T_z}{\partial t \partial z} + \frac{D_H}{e}\frac{\partial^3 T_x}{\partial z \partial x^2}. \tag{16}$$

Note that in the case of instantaneous thermalization ($T_z = T_x$) Eq. (16) has the following solution:

$$E_{z\,therm.} = -\frac{1}{e}\frac{\partial T_z}{\partial z}, \tag{17}$$

which coincides with the previous result from [15]. For clarity, Eq. (16) can be also written for the function $G = E_z + \frac{1}{e}\frac{\partial T_z}{\partial z}$, describing non-potential part of electric field, i.e. the difference between the full solution $E_z(x,z,t)$ and solution (17) for the case of instantaneous thermalization:

$$\frac{\partial G}{\partial t} - D_H \frac{\partial^2 G}{\partial x^2} = \frac{D_H}{e}\frac{\partial^3 (T_x - T_z)}{\partial z \partial x^2}. \tag{18}$$

The source in the right-hand side of Eq. (18) is proportional to the curl of electronic pressure, so the full solution for the electric field consists of the "potential" term $\sim \frac{\partial T_z}{\partial z}$ and "non-potential" part $G$. If we suppose that the optical field increases only the $z$-component of the temperature, then the function $E_z(x,z,t)$ has three basic timescales: duration of heating, isotropization time and the time of electron gas cooling.

It is also important that Eq. (16) allow to find the waveform of sub-THz electromagnetic pulse emitted from the metal surface (due to tangential electric field continuity at the boundary). From this point of view, developed theory describes the mechanism of optical-to-THz conversion, principally different from previously proposed ponderomotive [16, 17], drag-current [14, 18] and thermal [14, 15] mechanisms. Below we present results of numerical modeling and demonstrate that Eq. (16) gives a fairly good approximation for the electric field both inside the metal and at the boundary.

In the numerical scheme we solve the Cauchy problem for the system of modified hydrodynamic equations (3)–(4), together with Maxwell equations (7)–(8) and heat conductivity equations under the boundary:

$$\frac{\partial T_x}{\partial t} - D_T \frac{\partial^2 T_x}{\partial x^2} = \alpha(T_z - T_x), \tag{19}$$

$$\frac{\partial T_z}{\partial t} - D_T \frac{\partial^2 T_z}{\partial x^2} = \alpha(T_x - T_z) + (\mathbf{j}_{opt}, \mathbf{E}_{opt}), \tag{20}$$

where $\alpha$ is the isotropization rate. The term $(\mathbf{j}_{opt}, \mathbf{E}_{opt})$ is the absorption power written under assumption that the optical electric field is mostly parallel to the surface and so only the $z$-component of the temperature is initially increased. For simplicity, heat transfer to the crystal lattice is not included in Eqs. (19)–(20), but at the discussed timescales it has no significant influence on the studied electrodynamical processes. We consider impenetrable metal border, which leads to zero normal heat flux and zero normal velocity of electrons at the interface. The details of the numerical scheme can be found in [16], where similar simulations were used for investigation of the instant second-order nonlinear response of metal.

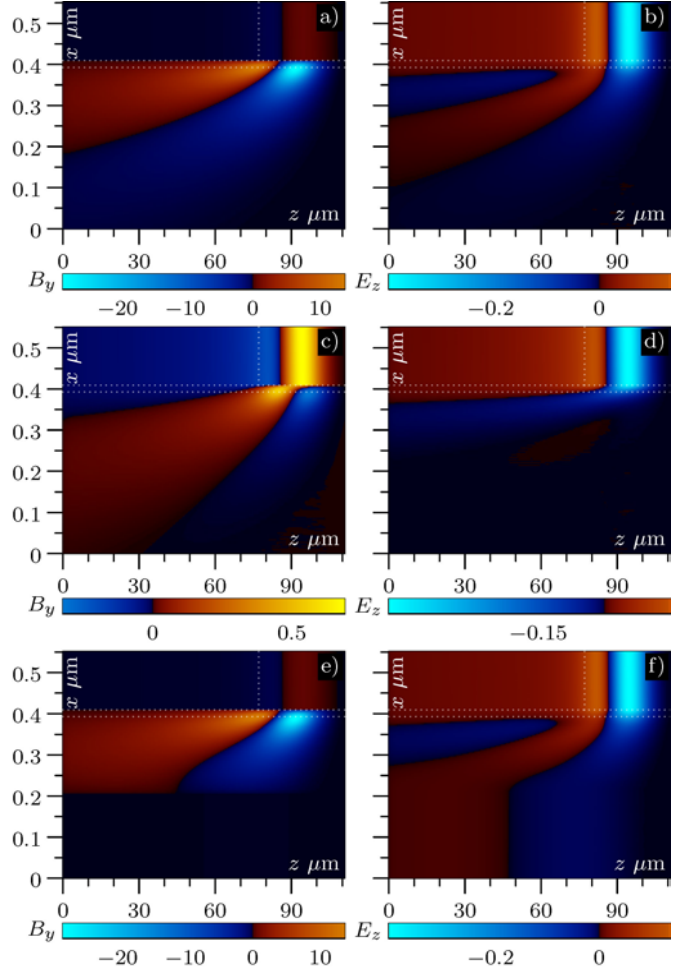

**Fig. 3.** Distributions of low-frequency magnetic field $B_y$ and tangential electric field $E_z$ inside the metal and near the boundary (in arb. units). Optical skin-layer is shown by horizontal dashed lines. a, b) Bulk metal, isotropization time 55 fs. c, d) Bulk metal, isotropization time 5.5 fs. e, f) Metal film with the thickness 180 nm, isotropization time 55 fs.

In the numerical modeling we use the following basic parameters: plasma frequency $\omega_p = 1.8 \cdot 10^{16}\ s^{-1}$, skin-layer depth 17 nm, collision frequency $\nu = 1.8 \cdot 10^{14}\ s^{-1}$, optical frequency $\omega = 2.5 \cdot 10^{15}\ s^{-1}$ (wavelength 800 nm), electronic thermal diffusivity $D_T = 50\ cm^2 s^{-1}$, magnetic diffusivity $D_H = 500\ cm^2 s^{-1}$, isotropization time $\alpha^{-1} = 55\ fs$ and laser pulse duration 40 fs (any differences are indicated in figures' captions). Distributions of electric and magnetic fields obtained in numerical modeling are shown in Fig. 3 for the cases of bulk metal (a–d) and metallic film (e, f). Penetration depth of low-frequency magnetic field is noticeably larger than the optical skin depth, which is a consequence of diffusive propagation of magnetic field inside the metal. Numerical results shown in Fig. 3 demonstrates that magnetic field inside the metal is up to ~50 times higher in the case of delayed isotropization (55 fs) in comparison with the case of almost instantaneous isotropization (5.5 fs). However, radiated fields have roughly the same amplitude (see Fig. 5 below). Note that in Fig. 3 and further the scales of *x* and *y* axis differs 200 times for better image of the internal structure of fields and currents at nanometer scales.

Distribution of bulk currents is shown in Fig. 4 (on the background of magnetic field). In both cases of semi-infinite medium and thin film, clear vortex structure of currents was obtained. Note that there are large regions of high vorticity with significant magnetic field without any charge density perturbation. According to numerical modeling data, maximal magnetic field is reached in the region located at a depth of 5 to 40 nm under the metal surface having a length of ~10 microns. In this Letter we study only non-magnetic metals with Drude-like electric properties and the investigation of magnetized medium dynamics is beyond our current consideration. However, discussed mechanism of magnetic field generation probably makes a significant contribution to ultrafast demagnetization of ferromagnetic films under the laser pulse action [19-22].

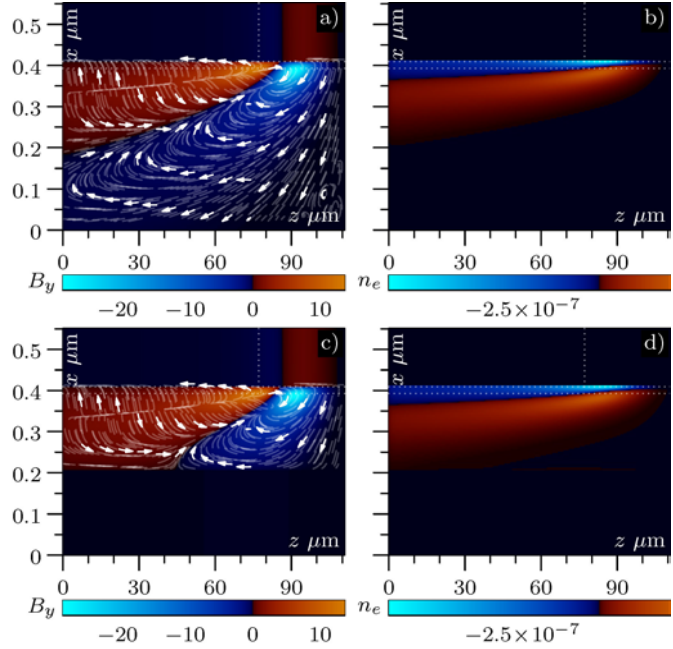

**Fig. 4.** Low-frequency bulk currents on the background of magnetic field $B_y$ in the case of (a) semi-infinite metal and metal foil (c). Panels (b) and (d) show corresponding electron density perturbation.

Comparison of the modeling results with the analytical theory developed in the first part of the Letter is shown in Fig. 5. We compare sub-THz electromagnetic signal radiated from the surface due to the discussed generation mechanism and electric field distributions inside the metal. One can see that the approximate diffusion-like equation (16) describe low-frequency electric field with fairly good accuracy. According to analytical estimations, when the difference of "temperatures" $T_x$ and $T_z$ is ~1 eV, the internal fields $E_x$ and $H_y$ reach values of several MV/cm and ~ 1 tesla respectively.

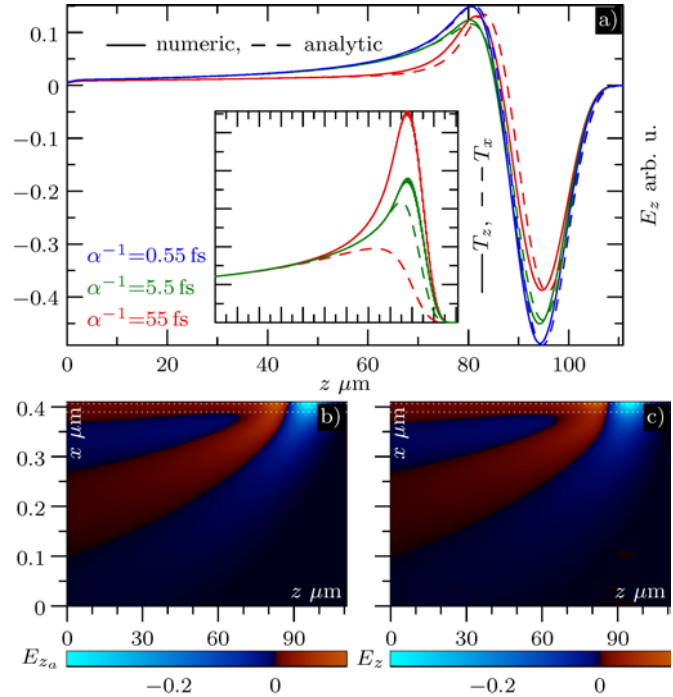

**Fig. 5.** Comparison of numerical modeling results with approximate analytical model. a) Sub-THz electric field above the surface of metal obtained in the framework of full hydrodynamical description (solid lines) and the solutions of approximate Eq. (16) (dashed lines). Inset: dynamics of distribution isotropization ($T_x$ – solid line, $T_z$ – dashed line). Lines of different colors correspond to different isotropization rates. b, c) Tangential electric field $E_z$ inside the metal according to Eq. (16) (b) and numerical modeling (c).

In conclusion, we found that anisotropy of the electronic distribution function inside the laser-excited metal becomes a non-potential force generating strong low-frequency magnetic field. Magnitude of generated internal field critically depends on the isotropization rate; it decreases by two orders when the isotropization time changes from 55 fs to 5.5 fs. Approximate analytical approach shows that magnetic field propagates deep into the metal in a diffusive way. We also obtained numerically full distributions of electromagnetic fields and bulk currents inside the metal, and demonstrated the presence of significant magnetic field at the depth up to 3-4 optical skin-layers. We believe that these theoretical results are promising for the interpretation of experimental data on optical-to-THz conversion, second harmonic generation, ultrafast demagnetization and other effects based on nonlinear electrodynamical properties of metals at subpicosecond timescales.

**Acknowledgements**. The work was financed by RFBR grant No 18-42-520023 r_a. Also I.V.O. is grateful for support to the Foundation for the Advancement of Theoretical Physics and Mathematics "BASIS".